\documentclass[twocolumn,showpacs,prb,amsmath,amssymb,floatfix]{revtex4}

\usepackage{graphicx}
\usepackage{subfigure}
\usepackage{dcolumn}
\usepackage{bm}

\begin{document}

\title{Magnetism in tunable quantum rings}

\author{G. B\aa rdsen}
\author{E. T{\"o}l{\"o}} 
\author{A. Harju}

\affiliation{ 
  Department of Applied Physics and Helsinki Institute of Physics,
  Helsinki University of Technology, P.O. Box 4100, FI-02015 HUT, Finland 
}

\date{\today}

\begin{abstract}
  We have studied the spin structure of circular four-electron quantum rings
  using tunable confinement potentials. The calculations were done using the
  exact diagonalization method. Our results indicate that ringlike systems can
  have oscillatory flips between ferromagnetic and antiferromagnetic behaviour as a function of the magnetic field. Furthermore, at constant external
  magnetic fields there were seen similar oscillatory changes between
  ferromagnetism and antiferromagnetism when the system parameters were
  changed. According to our results, the magnetism of quantum rings 
  could be tuned by system parameters.
\end{abstract}


\maketitle

\section{Introduction}

During the last two decades, there has been seen an increasing
scientific and technological interest in spin related phenomena and in
possible application of these in future data processing, communication
and storage.\cite{awsch,sarma2} Quantum dots have been proposed as
components in few-electron spintronics devices, such as spin filters
or spin memories,\cite{sarma2,recher} and in spin-based quantum
computation devices.\cite{loss}

Existing fabrication techniques allow construction of semiconductor quantum
rings of nanometer dimensions containing only a few
electrons.\cite{chring,lorke,hanson} Nanoscopic quantum rings are sufficiently
small systems to show quantum effects, and large enough to be able to trap
magnetic flux in their interior, when subjected to an experimentally reachable
magnetic field. The trapping of magnetic flux quanta gives rise to interesting
effects, such as persistent currents and other periodic properties related to
the Aharonov--Bohm effect.\cite{chring,ab}

The main focus of this paper is on the magnetic properties of the quantum
rings. The previous study by Koskinen et al.\cite{kosk2} for one type of
quantum rings has shown that a model of localized charges and
antiferromagnetic coupling of the nearest-neighbor spins, corresponding to an
antiferromagnetic Heisenberg model, captures the physics of systems they
study. This means, for example, that when the angular momentum of the system
is increased by making the magnetic field stronger, the change of the total
spin of the system follows values found using the antiferromagnetic Heisenberg
model.  On the other hand, a study of a hard-wall quantum dot\cite{hanc} where
strong electron-electron interaction forces the electron density to be
ring-shaped, has found a ferromagnetic behavior of the system as a function of
the magnetic field. Motivated by this discrepancy, we study tunable quantum
rings and aim to identify the underlying physics that leads to different
magnetic properties for the various quantum rings. 

Our results show interestingly that the same quantum ring system can show
oscillation between ferromagnetic and antiferromagnetic behavior as a function
of the magnetic field. A similar control of the magnetism can be obtained by
changing the width or the radius of the quantum ring.

\section{Model and computational Method}

We use an effective-mass approximation and model the semiconductor
quantum rings as two-dimensional systems with the Hamiltonian
\begin{equation} \label{eq:Hmicr} H=\sum_{i=1}^{N}\bigg(\frac{(-i\hbar
    \nabla_{i}+e\mathbf{A}_{i})^{2}}{2m^{*}}+V(r_{i})\bigg)+\frac{e^{2}}{4\pi
    \epsilon}\sum_{i<j}^{N}\frac{C}{r_{ij}},
\end{equation}
where $N$ is the number of electrons, $\mathbf{A}$ is the vector potential of
the perpendicular magnetic field $\mathbf{B}=B\mathbf{u}_{z}$, $V$ is the
external confinement potential, $m^{*}$ is the effective electron mass, $C$ is
the Coulomb constant (normally 1), and $\epsilon $ is the dielectric
constant. The electrons are restricted to the Cartesian $xy$ plane, and the
magnetic field is normal to the plane. In our calculations, we use the
effective parameter values of GaAs, namely, $m^{*} = $ 0.067$m_{e}$ and
$\epsilon = $ 12.7. To enhance the spin effects, we have left out the Zeeman
potential. This can be justified since experimentally the gyromagnetic factor,
and thereby also the Zeeman term, can be made vanishingly small.\cite{lead}

To control the spin effects, we use confinement potentials of the form
\begin{equation}
  V(r) = \frac{1}{2}m^{*}\omega_{0}^{2}r^{2} + V_{0}\exp(-r^{2}/\sigma^{2}),
\end{equation} 
where $\omega_{0}$ determines the confinement strength and $V_{0}$ determines
the strength of the Gaussian perturbation. We set $\hbar \omega_{0} = $ 5 meV
and $\sigma = $ 2 $a_{B}^{*}$, where $a_{B}^{*} \approx $ 10.03 nm is the
effective Bohr radius of GaAs.  By tuning the Gaussian perturbation stronger,
we can make the system more and more ring-shaped. In addition, we have used a
parabolic ring potential given by
\begin{equation}
  V(r) = \frac{1}{2}m^{*}\omega_{0}^{2}(r-r_{0})^{2} 
\label{ring2}
\end{equation} 
to model a narrow quantum ring.

In the exact diagonalization (ED) calculations, our single-particle states are
the one-body eigenstates of the Hamiltonian $H$. In cases when the confinement
potential $V$ is parabolic, the single-particle eigenstates are the
Fock-Darwin states.\cite{qmc} For systems with perturbed parabolic or
parabolic ring potentials, we calculate the single-electron states as linear
combinations of Fock-Darwin states, all with the same angular momentum,
corresponding to the 15 lowest energy values. The ED calculations take into
account Landau-level mixing, as the 25 lowest-lying eigenstates, irrespective
of Landau level, are included in the computational procedure. The ground state
eigenvalues of the Hamiltonian matrix of the ED method were obtained using the
Lanczos diagonalization method, and the interaction matrix elements were
calculated using numerical integration. The principles of the ED method are
described in Ref.~\onlinecite{sigga}.

When the electrons in a quantum ring become sufficiently localized, charge and
spin excitations separate from each other.\cite{kosk} This phenomenon is a
non-perturbative effect due to the strong correlations in quantum ring systems
at high magnetic fields. The qualitative behavior of the many-particle
spectrum of a quasi-one-dimensional system has been described by the lattice
Hamiltonian\cite{kosk2,vief,mare}
\begin{equation} \label{eq:effHam}
  H = J\sum_{i,j}\mathbf{S}_{i}\cdot
  \mathbf{S}_{j}+\frac{1}{2I}L^{2}+\sum_{\alpha }\hbar \omega_{\alpha
  }n_{\alpha },
\end{equation}   
where the first term is a Heisenberg Hamiltonian that models the spin degrees
of freedom, the second term is a rigid rotation of the system, and the last
term includes vibrational modes of the localized electrons. The parameters of
the model are the nearest-neighbor coupling constant between the spins $J$,
the total moment of inertia $I$, and the vibration frequency $\omega_{\alpha
}$. The angular momentum of the ring is given by $L$ and $n_{\alpha }$ is the
number of excitation quanta of the vibrational mode.  In an antiferromagnetic
system the coupling is such that $J > $ 0, and in a ferromagnetic system $J <
$ 0. 

The total angular momentum--spin pairs of the lowest eigenstates for the
effective Hamiltonian (\ref{eq:effHam}) can be calculated using exact
diagonalization, or using group-theoretical methods.\cite{kosk2} In
Table~\ref{tab:af}, we have given the angular momentum and spin states of the
ferromagnetic and antiferromagnetic systems of four particles, as calculated
by the exact diagonalization. The spin sequence is $N$-periodic as a function
of the angular momentum, where $N$ is the number of electrons. One can see
that for four particles, the antiferromagnetic behavior is manifested by the
three consecutive ground states with spin equal to one, and one signature of
the ferromagnetic model is the occurrence of ground state with spin equal to
two. One should also note that there is one ground state with total spin equal
to zero for both versions of the Heisenberg model.

\begin{table}
  \caption{Total angular momenta $L$ and their corresponding spin values $S$
    for the ferromagnetic and antiferromagnetic ground states of the
    Heisenberg model of a four-electron ring. Spin states corresponding to
    higher angular momenta are obtained by utilizing the periodicity of the
    spin sequence.}
  \begin{ruledtabular}
    \label{tab:af}
    \begin{tabular}{r c c c c c}
      $L$ &  & 0 & 1 & 2 & 3 \\
      \hline
      $S$ & Ferromagnetic & 0 & 1 & 2 & 1 \\
      $S$ & Antiferromagnetic & 1 & 1 & 0 & 1 \\
    \end{tabular}
  \end{ruledtabular}
\end{table}

\section{Results}

\subsection{Gaussian perturbation ring}

We start with the results for the parabolic dot that has been perturbed with
the Gaussian potential in the center to form a quantum ring. 
\begin{figure}
\includegraphics[width=.99\columnwidth]{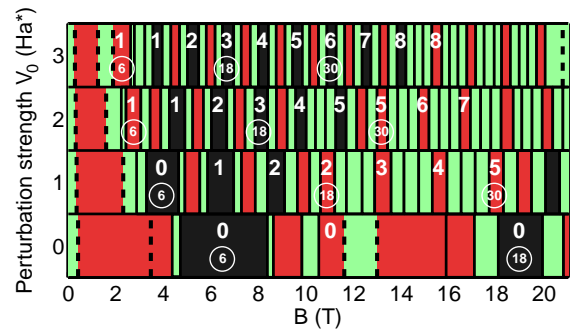}
\caption{(Color online) Ground state spin polarization as a function
  of the magnetic field $B$ for a four-electron parabolic quantum dot
  with Gaussian perturbations of different strengths $V_{0}$. The spin
  states are 0 (red/middle grey), 1 (green/light grey), and 2
  (dark grey). The upper integers denote the number of central
  vortices found in each of the different states, and the integers
  surrounded by a ring denote the corresponding angular momenta. All
  states at zero magnetic field have angular momentum 0. If two
  subsequent states are separated by a solid line, the angular
  momentum increases by one when moving to the state at right, and if
  they are separated by a dashed line, the angular momentum increases
  by two. } \label{fig:phaseD}
\end{figure}
In Fig. \ref{fig:phaseD}, we have plotted spin ground states as a function of
the magnetic field $B$ for four-electron perturbed parabolic dots with
perturbation strengths $V_{0} = $ 0, 1, 2, and 3 Ha$^{*}$ (1 Ha$^{*} \approx $
11.3036 meV). For all these perturbation strengths, the radius of the quantum
ring is around 2 $a_{B}^{*}$. For $V_0=0$, corresponding to a pure
parabolic quantum dot, the states with $L=6$ and 18 are the only fully
spin-polarized states (dark gray regions in phase diagram), as also found in
Ref.~\onlinecite{kosk} in the lowest Landau level approximation.  The state
with $L=6$ is called the maximum-density droplet (MDD),\cite{macd}
corresponding to the fractional quantum Hall states with filling fraction
$\nu=1$, and the state with $L=18$ corresponds to $\nu=1/3$.\cite{laughlin1}
Between these two states the total spin has values of zero and one, but these
do not follow the predictions from the antiferromagnetic Heisenberg model,
most clearly seen from the fact that there are no regions with three
consecutive ground states with spin equal to one. This is not surprising, as
we expect the Heisenberg model to be relevant only in the limit where
electrons are strongly in a ring-shaped confinement.

Next we turn on the Gaussian perturbation, and one can see that the
spin-polarized state with $L=6$ at $V_0=0$ splits into several spin-polarized states with
different angular momentum. In addition, between these $S=2$ states, the total
spins follow the ferromagnetic Heisenberg spins given in
Table~\ref{tab:af}. The only exception on this rule is found at $V_0=2$ around
$B=11$~T, where $S=1$ is found instead 0.

Outside these ferromagnetic regions at stronger magnetic fields, one can see
antiferromagnetic behavior. The antiferromagnetic phase can be identified by the three $S=1$ states
in a row, separated by the $S=0$ state. In addition, antiferromagnetic
behavior could develop at weak magnetic fields, as at $V_0=3$ there are
already three consecutive states with $S=1$. As will be shown later in this paper, when the ring is sufficiently narrow there is indeed found an antiferromagnetic region at magnetic fields lower than the first ferromagnetic region.

In a previous study\cite{hanc} of a hard-wall quantum dot with a radius $R = $
5 $a_{B}^{*}$ and maximum of the electron density around 3 $a_{B}^{*}$, the
spin structure was ferromagnetic for angular momenta $L$ between 6 and
18. Such a magnetic structure is almost the same as that found for the
perturbed parabolic system with perturbations $V_{0} = $ 1 and 2. The magnetic
structure of the wide quantum ring is thus for angular momenta 6 $\leq L \leq
$ 18 similar to that of a hard-wall quantum dot.

\subsection{Vortex structure}

From the phase diagram shown in Fig. \ref{fig:phaseD}, one can also see that
the state with angular momentum $L=6$ changes spin when the Gaussian impurity
is made stronger.  This change and many other details of the phase diagram can
be understood by studying the vortex structure of the many-body wave function.

\begin{figure}
  \includegraphics[width=\columnwidth]{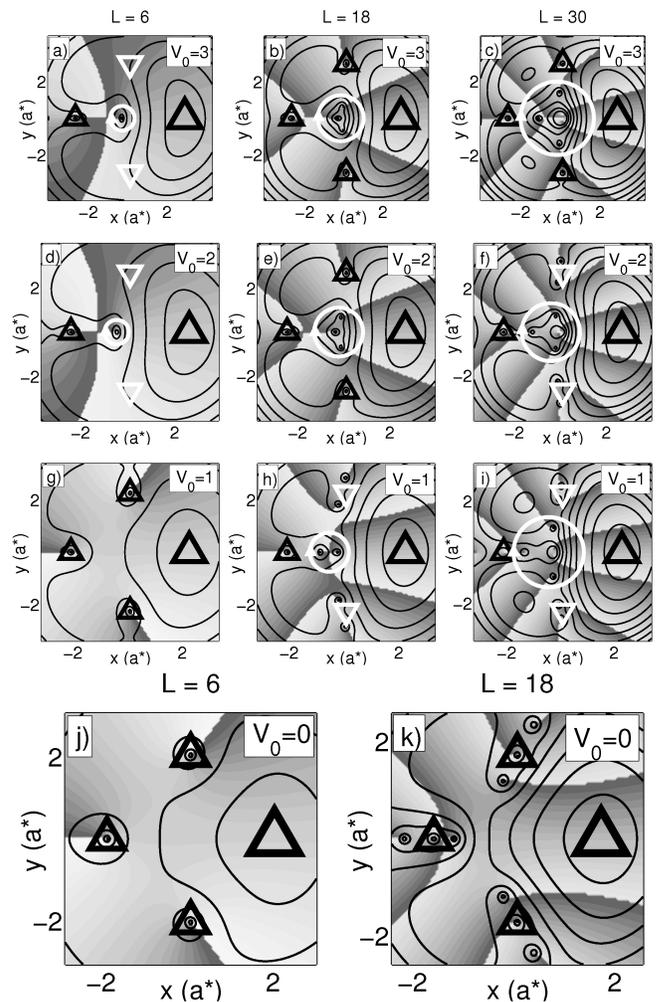}
  \caption{Conditional wave functions consisting of the charge
    density, plotted with contour lines (logarithmic scale), and the
    phase of the wave function, illustrated by the gray-scale
    shading. The fixed spin up electrons are marked with black
    triangles pointing up, and the spin down electrons with white
    triangles pointing down. The probe particle is marked with a
    larger triangle. The clusters consisting of central vortices
    (nodes), which can be considered as giant vortices, are denoted by
    circular arrows. 
} \label{fig:condWb}
\end{figure}

When subjected to an external magnetic field, the electrons in a quantum dot
are forced to rotate. If the rotation is sufficiently strong, vortices are
formed in the electron liquid.\cite{vortex} One can consider the vortices as
quasiparticles that are holes in the occupied Fermi sea.\cite{mare} The
vortices are seen as zero points in the conditional electron density, where
the phase of the wave function changes as a multiple of 2$\pi $ for a path
enclosing the node. As shown in the fractional quantum Hall effect (FQHE)
theory,\cite{eza} the formation of vortices is a result of quantization of the
magnetic field, and each vortex is associated with an integer number of Dirac
flux quanta.

To analyze the nodal structure of the perturbed quantum dot, we have
calculated conditional wave functions,\cite{vortex} defined for an
$N$-particle system as
\begin{equation}
  \psi_{c}(\mathbf{r}) = \frac{\Psi(\mathbf{r},\mathbf{r}^{*}_{2},\dots
    ,\mathbf{r}^{*}_{N})}{\Psi(\mathbf{r}^{*}_{1},\mathbf{r}^{*}_{2},\dots
    ,\mathbf{r}^{*}_{N})},
\end{equation}
where $\mathbf{r}$ is the position of the moved particle 1, and
$\mathbf{r}_{i}^{*}$ is the most probable position of particle $i$ that are
found by maximizing the total electron density $|\Psi |^{2}$. The phase
$\theta $ is obtained from the relation
$\psi_{c}(\mathbf{r})=|\psi_{c}(\mathbf{r})|\exp(i\theta(\mathbf{r}))$ and the
conditional electron density is defined as $|\psi_{c}(\mathbf{r})|^{2}$.

Now returning to the question related to the spin of the $L=6$ state, a
starting point for this analysis is the conditional wave function of the
parabolic dot shown in Fig. \ref{fig:condWb}(j). In this, one can see a Pauli
vortex on top of each electron, except the probe particle on right. The
vortices are shown by the discontinuous jumps in the gray-scale from white to
black when the electrons are circulated in a clockwise fashion.  The fact that
the vortex number is the same as the electron number shows that the state is a
finite-size example of a quantum Hall state with filling fraction $\nu=1$.

For $V_0=1$, shown in Fig. \ref{fig:condWb}(g), the spin is still the same and
the conditional wave function is nearly identical to the $V_0=0$
state. However, for $V_0=2$ the spin has changed from 2 to 0, and the
conditional wave function in Fig. \ref{fig:condWb}(d) looks completely
different. There is still one Pauli vortex on top of the left-most electron,
but the two other electrons have switched spins. In addition, there is now one
vortex that is located at the center of the system. It turns out that one
needs to combine two Pauli vortices in order to make one vortex at the center
of the dot. From this data, one can understand the change in the total spin as
follows: The system places the vortices so that it minimizes the total
energy. Without the Gaussian impurity at the center of the dot, it is
energetically favorable to place the vortices on top of the electrons to
reduce the Coulomb repulsion. When the potential at the center of the
dot is raised by the Gaussian impurity, at some strength it is more favorable
to reduce the probability of the electrons to be at the center of the dot by
placing a vortex there.

One can do a similar analysis for the $L=18$ state, which for $V_0=0$
corresponds to the $\nu=1/3$ state. The correspondence can be seen by the
three vortices bound to each electron, as shown in
Fig. \ref{fig:condWb}(k). The reason for the small separation of the vortices
is due to long-range nature of the interaction.  For $V_0=1$ shown in
Fig. \ref{fig:condWb}(h), there are already two vortices at the center of the
dot, and to create these, four Pauli vortices have vanished from the
system. Due to this, the spin has again changed, and the opposite spins have
only two vortices bound to them. As a general rule, opposite spin electrons
can have even vortices bound to them, and the antisymmetry requirement of same
spin electrons forces the vortex number to be odd. The vortex structure of the
$S=0$ states is similar to that of a $\nu = 2/3$ Halperin
state,\cite{cp,persp,halp} apart from the vortices at the center of the dot.
Going to $V_0=2$ shown in Fig. \ref{fig:condWb}(e), a third vortex has
appeared at the center, and now again the total spin of the system has
changed. Finally, at $V_0=3$ shown in Fig. \ref{fig:condWb}(b) the conditional
wave function is nearly identical to the $V_0=2$ case.

The same trend can be found at even larger values of the angular momentum, and
as a final example, we show the data for the $L=30$ states in
Figs. \ref{fig:condWb}(c),(f), and (i). The main difference to the previous
examples is that when the number of central vortices grows, also the area
they occupy seems to be larger.

Analyzing the vortex structure of the other states in detail enables us to
label the central vortex numbers of the ground states. These numbers are given
in Fig. \ref{fig:phaseD} for the most interesting states. One can see that in
general, the number of central vortices grows as a function of the magnetic
field. However, at each fixed Gaussian impurity strength, there is a point
where it is energetically favorable to add Pauli vortices instead of the
central vortices, and at this point the number of central vortices is constant
although the angular momentum increases. At these points, the ferromagnetic
behavior is changed to antiferromagnetic. Somewhat similar transitions are
seen at the magnetic fields below the first ferromagnetic states.

It is also interesting to analyze the ground states at magnetic fields around
6--7~T. At $V_0=0$, the ground state is the $S=2$ state with $L=6$ corresponding to
$\nu=1$. When the Gaussian impurity is made stronger, there are more and more
vortices at the center of the system, and the angular momentum is increased.
However, the spin of the ground state does not change. This shows that the
$\nu=1$ state is in some sense stable even for this small particle number when
one pierces it with three fluxes at the center. By this we mean that the
vortex structure of the conditional wave function is the same apart from the
central vortices, see Figs.~\ref{fig:phaseD}(b) and (j).

\subsection{Narrow quantum ring}

Based on the analysis presented above, one would expect to find an
antiferromagnetic region at low magnetic fields, if the ring is made
sufficiently narrow (see the region with $B = $ 2--4 T and $V_{0} = $ 3 Ha$^{*}$ of Fig. 1). To investigate this, we switch the confinement potential
to
\begin{equation}
  V(r) = \frac{1}{2}m^{*}\omega_{0}^{2}(r-r_{0})^{2},
\end{equation} 
where the confinement strength is taken to be 40 meV and the radius
8$a_{B}^{*}/\pi\approx 2.5 a_{B}^{*}$.  This system is a very narrow ring with
approximately the same radius as the ring studied above.

\begin{figure*}
  \includegraphics[width=.65\columnwidth]{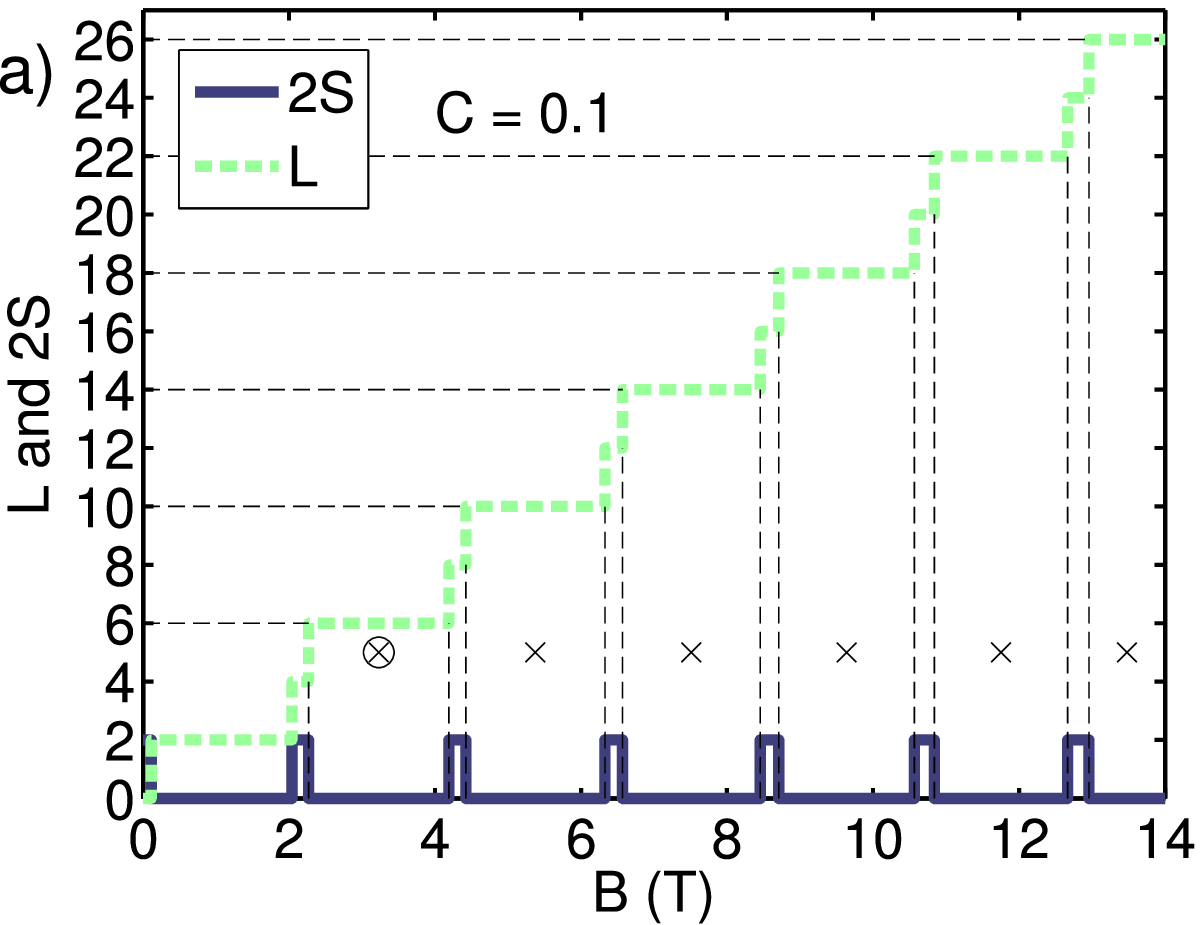}
  \includegraphics[width=.65\columnwidth]{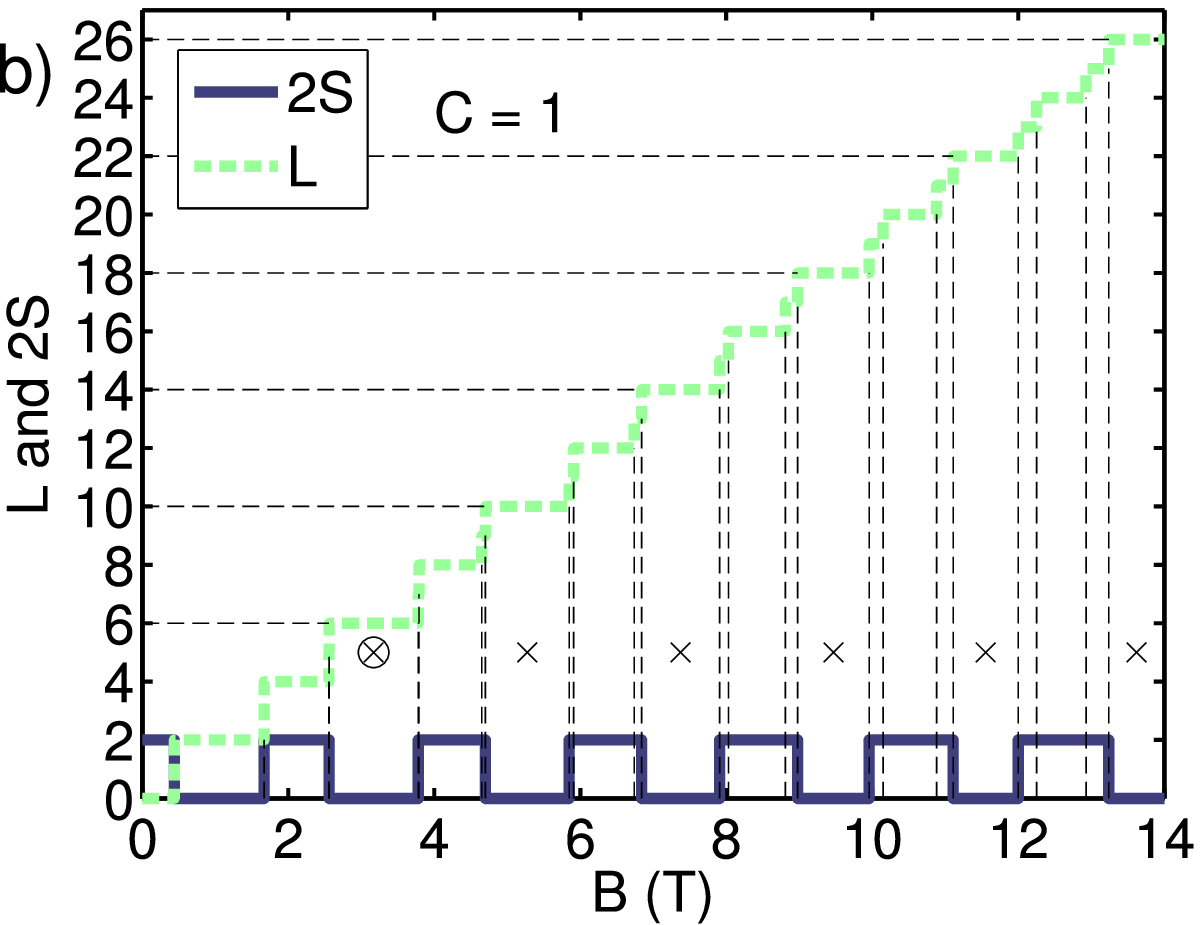}
  \includegraphics[width=.65\columnwidth]{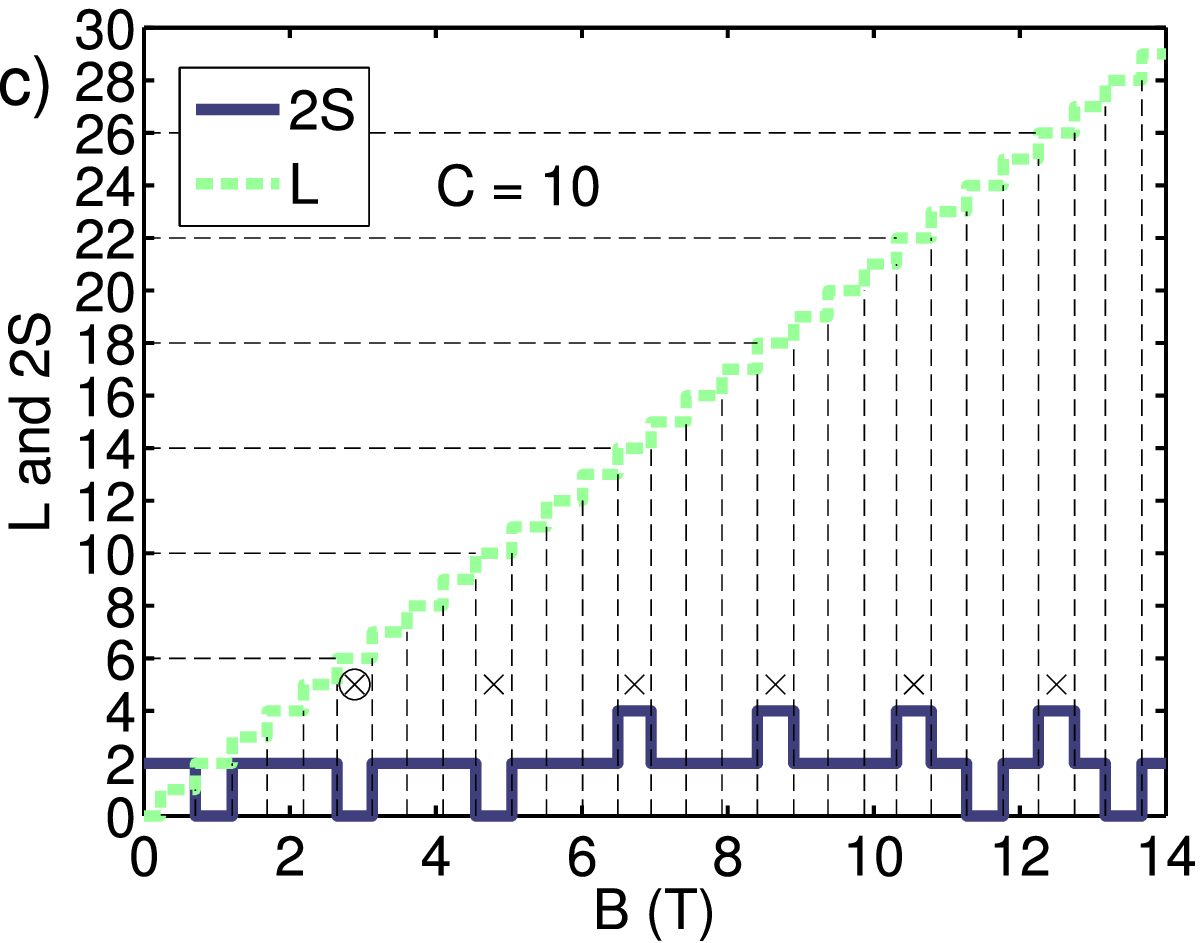}
  \caption{(Color online) Total angular momentum $L$ and spin $S$ of the ground
    states given as functions of the magnetic field $B$ for a
    four-electron parabolic ring with 40 meV confinement strength and
    radius 8$a_{B}^{*}$/$\pi $, where $a_{B}^{*}$ is the effective
    Bohr radius. The Coulomb constant is (a) $C = $ 0.1, (b)
    $C = $ 1, and (c) $C =$ 10. The crosses denote states with
    magic angular momenta $L_{k} = L_{\textrm{MDD}} + kN$, where $k
    \in \{0,1,2,\dots \}$, $L_{\textrm{MDD}} = (N-1)N/2$ is the
    angular momentum of the maximum-density droplet (MDD), and $N$ is
    the number of electrons. The state with angular momentum
    $L_{\textrm{MDD}}$ is marked with a cross surrounded by a
    circle.} \label{fig:ring}
\end{figure*}

In Fig. \ref{fig:ring}, the total angular momentum and spin of the ground
state is plotted for the narrow ring.  We have varied the Coulomb constant
between the values $C = $ 0.1, 1, and 10. In reality, the interaction strength
can be changed by changing the quantum ring radius.\cite{sigga} The case with
$C=1$, shown in Fig. \ref{fig:ring}(b), corresponds to the natural interaction
strength, and serves as basis for the comparison with the ring used above. For
this case, the total spin has values zero and one, and above $B=4$ T there are
three states with spin one between each ground state with spin equal to zero,
in agreement with the antiferromagnetic Heisenberg model. Thus, as expected there is an antiferromagnetic region at low magnetic fields when the ring is sufficiently narrow.

When the Coulomb interaction is made weak by setting $C=0.1$, two of the three
consecutive $S=1$ states are no longer ground states as shown in
Fig.~\ref{fig:ring}(a). Now, although the spin has values 0 and 1, the system
does not behave as the antiferromagnetic model. This shows that the
localization induced by the strong interaction is a necessity for
antiferromagnetism.

On the other hand, when we make the interaction stronger by setting $C=10$,
one can see that fully spin polarized ground states are found. In addition,
the magnetic field range of each ground state is nearly the same.  In the magnetic field region between 6 and 10 T, the ferromagnetism is not as pure as the antiferromagnetism for the $C=1$ case, as now the ground states with $S=0$ are missing. Also, the system is antiferromagnetic upto $B\approx 6$ T. We conclude that a transition between antiferromagnetic and ferromagnetic behaviour also can be controlled by the Coulomb interaction strength, or physically by tuning the radius of the quantum ring. 

\section{Conclusion}

Our results indicate that both ferro- and antiferromagnetic phases are found
for quantum rings in a strong magnetic field. In addition, the occurrence of the different magnetic phases are not trivial to predict, as the magnetic phase is determined by the magnetic field region, the ring width and the radius of the system. We found that the magnetic phase changes oscillatory as a function of the different system parameters. Knowledge of the nontrivial magnetic phase structure could be very fruitful for experiments, as the observed phenomena opens the possibility for tunability of the magnetism by changing system parameters. For example, in an experiment, a metallic electrode with adjustable voltage could be used to achieve the control we had in our calculations when using the Gaussian impurity. \\

\begin{acknowledgments}
This study has been supported by the Academy of Finland through its Centers of
Excellence Program (2006--2011). ET acknowledges financial support from the
Vilho, Yrj{\"o}, and Kalle V{\"a}is{\"a}l{\"a} Foundation of the Finnish
Academy of Science and Letters. We also thank Henri Saarikoski for useful
discussions. 
\end{acknowledgments}

\end{document}